\documentclass[onecolumn,aps,floatfix,10pt,notitlepage,groupedaddress]{revtex4-2}
\pdfoutput=1
\usepackage{hyperref}
\usepackage{amsmath}
\usepackage{amsfonts}
\usepackage{graphicx}
\usepackage{matlab-prettifier}
\usepackage[none]{hyphenat}
\hypersetup{colorlinks, allcolors={blue}}
\newcommand{\fref} [1]{Fig.~\ref{#1}}
\newcommand{\Fref} [1]{Figure~\ref{#1}}
\newcommand{\ffref}[1]{Figs.~\ref{#1}}
\newcommand{\FFref}[1]{Figures~\ref{#1}}
\newcommand{\eref} [1]{Eq.~\eqref{#1}}
\newcommand{\cref} [1]{ref.~[\onlinecite{#1}]}
\newcommand{\sref} [1]{Sec.~\ref{#1}}
\newcommand{\aref} [1]{App.~\ref{#1}}
\newcommand{\Frac}{\displaystyle\frac}

\begin{document}

\title{Magic-angle twisted bilayer graphene under orthogonal and in-plane magnetic fields}
\author{Ga\"{e}lle Bigeard and Alessandro Cresti}
\address{\mbox{Univ. Grenoble Alpes, Univ. Savoie Mont Blanc, CNRS, Grenoble INP, CROMA, 38000 Grenoble, France}}

\begin{abstract}
We investigate the effect of a magnetic field on the band structure of bilayer graphene with a magic twist angle of 1.08$^\circ$.  
The coupling of a \mbox{tight-binding} model and the Peierls phase allows the calculation of the energy bands of periodic \mbox{two-dimensional} systems. 
For an orthogonal magnetic field, the Landau levels are dispersive, particularly for magnetic lengths comparable to or larger than the twisted bilayer cell size. 
A high \mbox{in-plane} magnetic field modifies the \mbox{low-energy} bands and gap, which we demonstrate to be a direct consequence of the minimal coupling.
\\[5mm]
\noindent \emph{This is the self-archived version of the paper:} \\
\hspace*{1cm}  Ga\"elle Bigeard and Alessandro Cresti 2024 {\it J. Phys.: Condens. Matter} {\bf 36} 325502\\
\hspace*{1cm}  doi: \href{\doibase 10.1088/1361-648X/ad4431}{10.1088/1361-648X/ad4431}
\\[3mm]
This is the version of the article before peer review or editing, as submitted by an author  to Journal of Physics: Condensed Matter. IOP Publishing Ltd is not responsible for any errors or omissions in this version of the manuscript or any version derived from it. The Version of Record is available online at \href{\doibase 10.1088/1361-648X/ad4431}{10.1088/1361-648X/ad4431}.
\end{abstract}

\maketitle

\section{Introduction}
At magic angles, twisted bilayer graphenes (TBGs) exhibit almost flat \mbox{low-energy} bands within a small gap, as theoretically predicted~\cite{Santos2007,Morell2010,Bistritzer2011a} and experimentally observed~\cite{Jiang2019,Choi2019,Tomarken2019,Utama2020}. 
This has attracted much attention because of the resulting topological effects~\cite{Song2019,Liu2019,Nuckolls2020,Choi2021,HerzogArbeitman2022,Khalifa2023} and enhanced \mbox{electron-electron} coupling~\cite{Kerelsky2019,Choi2019,Burg2019} with interesting consequences~\cite{Andrei2020}, such as superconductivity~\cite{Cao2018,Lu2019,Yankowitz2019,Park2021}, ferromagnetism~\cite{Sharpe2019}, large \mbox{in-plane} orbital magnetization in the presence of strain~\cite{Antebi2022} and hysteretic anomalous Hall effect~\cite{Grover2022}.
In the literature, the band structure of TBGs in the presence of a magnetic field reveals interesting features, even in the absence of \mbox{many-body} interactions. 
For a magnetic field orthogonal to TBGs, many contributions focused on the Hofstadter butterfly~\cite{Moon2012,Hasegawa2013,Hejazi2019,Lian2020,Zhang2019}. 
From experiments on quantum Hall effect~\cite{Lee2011,SanchezYamagishi2012,Wu2021}, \mbox{low-energy} states turn out to be localized due to random spatial inhomogeneities of the interlayer coupling~\cite{Lee2011}. 
For \mbox{higher-quality} systems, such a localization can likewise be explained in terms of the results illustrated below. 
\mbox{Tight-binding} (TB) models and Peierls phase were used to calculate the energy of Landau levels (LLs)~\cite{Do2022} and the spatial distribution of states in TBG flakes~\cite{Landgraf2013}. 
The effect of an orthogonal magnetic field was investigated~\cite{Wang2012}, within the TB description and for not necessarily commensurate twist angles, by considering a large isolated cell in order to avoid edge effects. 
Bilayer graphene with AA or AB stacking and TBGs with an \mbox{in-plane} magnetic field have been less investigated, either by using the continuum approximation~\cite{Gail2011,Roy2013,Donck2016,Kheirabadi2016,Stauber2018a,Kwan2020} or TB models~\cite{Stauber2018,Kammermeier2019,Kheirabadi2022}. 
Still with an \mbox{in-plane} magnetic field, other interesting experimental and theoretical results mostly concern \mbox{low-temperature} superconducting states \cite{Cao2021a,Yu2021,Qin2021} at low magnetic field.  

In this work, we study the effect of a magnetic field on the band structure of a TBG with a magic twist angle of 1.08$^{\circ}$ and \mbox{non-interacting} electrons described by a TB Hamiltonian. 
Although continuum effective models~\cite{Tarnopolsky2019,Carr2019,Carr2019a,Koshino2020,Cao2021} are computationally convenient for treating extremely large unit cells corresponding to very small twist angles, they are less accurate in terms of microscopic description~\cite{Po2019}.  
We consider a \mbox{two-dimensional} periodic TBG, for which the Bloch theorem can be applied, although only commensurate twist angles can be considered and only \mbox{low-energy} bands can be calculated due to the large number of atoms in the primitive cell. 
This allows us to obtain the energy bands and spatial density probability of the eigenstates in the whole \mbox{two-dimensional} Brillouin zone. 
For an orthogonal magnetic field, we will deepen the analysis of the interplay between the magnetic length and the dispersive nature of the LLs, thus clarifying the different responses at relatively low and high magnetic fields. 
This will be done by scrutinizing the real-space profiles of Landau states in the TBG regions of different local stacking. 
Finally, we will focus on the effect of an \mbox{in-plane} magnetic field on the energy bands of a TBG, which we will demonstrate to be significant for small twist angles and a large magnetic field. 

As mentioned above, we use a non-interacting electron TB Hamiltonian. 
Although electron interactions may be relevant for very \mbox{high-quality} systems with almost flat bands, our \mbox{non-interacting-electron} calculations provide interesting results for \mbox{non-extremely} \mbox{high-quality} systems and represent a starting point for more advanced models. 
Moreover, the adopted \mbox{tight-binding} Hamiltonian model can serve as a basis for simulating quantum electron transport for example in the presence of disorder or more complex geometries, as Hall bars. 

\section{Methods}
Among the TB models for TBGs existing in the literature~\cite{Laissardiere2012,Fang2016,Lin2018,Pathak2022}, we adopt the one given in \cref{Lin2018}, because it properly reproduces the energy bands obtained by density functional theory for AB stacking, AA stacking and magic twist angle $\theta=1.08^{\circ}$, without having to relax the geometry~\cite{Uchida2014,Cantele2020}. 
This model limits the \mbox{in-plane} coupling to first neighbors, with value $t_\parallel$, while the \mbox{out-of-plane} coupling is given by
\begin{equation} \label{eq:PRB98_interlayer}
	t_\perp(r) \ = \ \alpha \ \Frac{\Delta z^2}{r^2 \ + \ \Delta z^2} \ \exp\left(-\Frac{\sqrt{r^2 \ + \ \Delta z^2} \ - \ \Delta z}{\lambda}  \right) \ ,
\end{equation}
where $r$ is the distance between two carbon atoms, $\Delta z$ is the \mbox{inter-layer} distance and $t_\perp(r)$ is set to zero beyond the \mbox{cut-off} distance 0.5~nm.
Table~\ref{tab:PRB98} reports the value of the parameters. 
\begin{table}[b!]
	\centering
		\begin{tabular}{| c | c | c | c |}
		\hline $\Delta z$    & $t_\parallel$ & $\alpha$ & $\lambda$ \\
		\hline 0.335~nm & 3.09~eV       & 0.39~eV         & 0.027~nm \\ \hline 
		\end{tabular}
	\caption{Parameters for the adopted TB model of \cref{Lin2018}.}
	\label{tab:PRB98}
\end{table}
In the presence of a magnetic field $\mathbf{B}$, the Hamiltonian is modified by the minimal coupling~\cite{Jackson1998} of the momentum operator
\begin{equation} \label{eq:minimal_coupling}
	\hat{\mathbf{p}}	\ \rightarrow \hat{\mathbf{p}}	\ - \ \Frac{e}{c} \mathbf{A}(\hat{\mathbf{r}})	 \ ,
\end{equation}
where $\mathbf{A}(\mathbf{r})$ is the vector potential, $\nabla\times\mathbf{A}=\mathbf{B}$, and $\hat{\mathbf{r}}$ is the spatial position operator.
For a TB Hamiltonian, \eref{eq:minimal_coupling} translates into the Peierls phase approximation~\cite{Peierls1933}.
In \mbox{two-dimensional} periodic systems~\cite{Cresti2021}, the flux of the magnetic field through the unit cell can only be a multiple of the quantum flux $\Phi_0 = h/(ec) \approx 4.136\times 10^3$~T~nm$^2$. 
As a consequence, the orthogonal component of the magnetic field must be multiple of a value that depends on the cell surface, contrarily to the \mbox{in-plane} component of the magnetic field, which flux is null.
We do not include here the effect of the Zeeman splitting~\cite{Zhang2006}, which, in the absence of \mbox{electron-electron} coupling and for negligible \mbox{spin-orbit} coupling in graphene~\cite{Guinea2010}, just lifts the degeneracy by $\Delta=g\mu_B |\mathbf{B}|\approx 0.12~B$[T]~meV~\cite{Sarma2011}.

\section{Results}
In \sref{sub:AA_AB}, we briefly review the LLs in AA and AB stacked bilayer graphenes. 
The physics behind the LLs in untwisted bilayer graphene bands, which is \mbox{well-known} in the literature, is crucial to explaining that behind TBGs. 
The reason is that, depending on the twist angle, TBGs show small to large subregions with roughly regular AA or AB stacking.
In \sref{sub:ORT}, we study the effect of an orthogonal magnetic field on a TBG at the magic angle $\theta=1.08^\circ$. 
The LL dispersion is found to be related, at small magnetic length, to that of untwisted bilayer graphene.
To better understand this relationship, we show the \mbox{real-space} profiles of the corresponding eigenstates. 
Depending on the specific magnetic field and energy, these states can be more or less localized on the AA or AB stacked subregions.
Due to the numerical burden and the unit cell size, our minimum magnetic field is about 28.36~T. 
Such a magnetic field is quite large for typical laboratories, which usually operate below 12~T. 
However, for the considered TBG, such a value still corresponds to a magnetic length larger than the primitive cell size. 
Therefore, we expect a completely similar behavior even at smaller magnetic fields, which are easily obtained experimentally. 
The higher magnetic fields at which some of the phenomena are observed are nevertheless within the reach of high magnetic field laboratories. 
Finally, in \sref{sub:PLANE}, we study the effect of an in-plane magnetic field on our TBG and provide a physical explanation. 
In this case, important effects such as band gap closure are visible at high magnetic fields.
We also focus on the special case of an AA graphene bilayer, which is equivalent to a zero twist angle. 
In this case, a few meV band gap opens at low magnetic fields, and is proportional to the magnetic field itself. 

\begin{figure}[b!]
	\centering	
		\includegraphics[width = 8 cm]{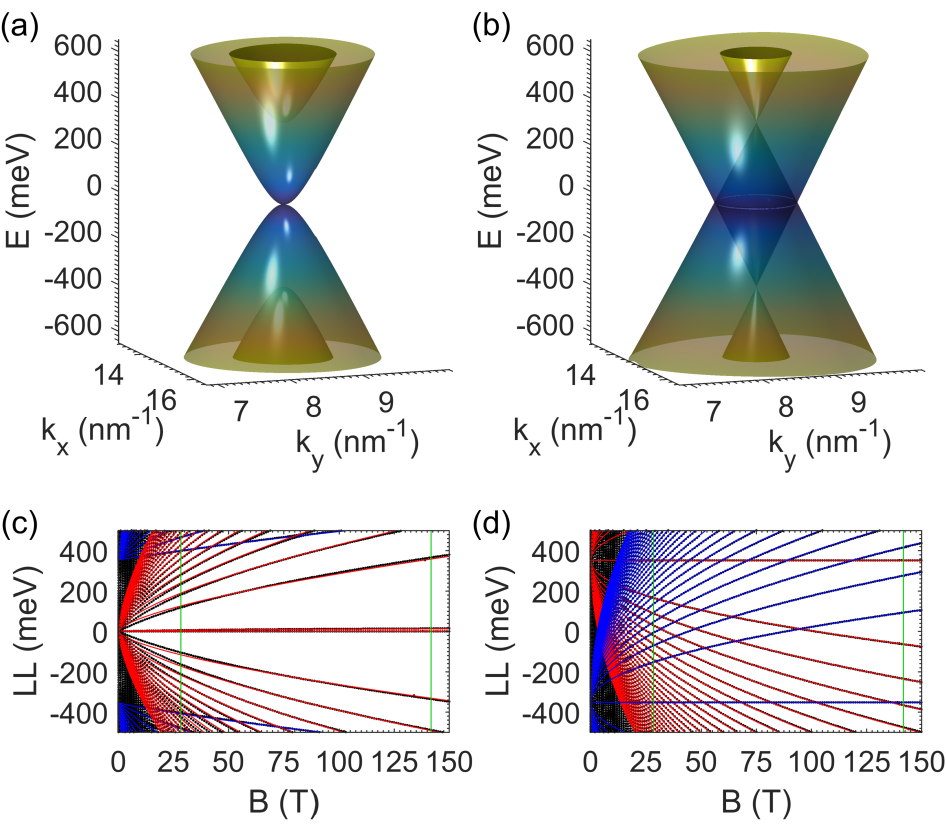}
	\caption{
	          \mbox{Low-energy} bands for bilayer graphene with AB stacking (a) and AA stacking (b) around the $K$ point, and energy of the resulting flat LLs for AB stacking (c) and AA stacking (d) in the presence of an orthogonal magnetic field $B$. 
	          The continuous lines in panels (c,d) correspond to an approximate fitting. 
						The vertical green lines correspond to values of the magnetic field considered in the paper, see \fref{fig:B_perp}, and are compatible with the TBG cell size at the magic twist angle $\theta=1.08^\circ$.}
	\label{fig:AB_AA}
\end{figure}

\subsection{Orthogonal magnetic field in bilayer graphene with AA and AB stacking\label{sub:AA_AB}}
Let us start by illustrating the effect of an orthogonal magnetic field on the energy bands of bilayer graphene with AB and AA stackings~\cite{Do2022}. 
\FFref{fig:AB_AA}(a,b) show the \mbox{low-energy} band structure around the $K$ point of the Brillouin zone and without a magnetic field.
For AB stacking, the bands display a \mbox{parabolic-like} dispersion and a semimetallic behavior, with two inner bands touching at $E=0$ and two outer bands starting at about $E=\pm 352$~meV~\cite{Laissardiere2012}. 
The same energy shift is observed for the AA stacking, which, however, preserves the linear dispersion of monolayer graphene, with Dirac cones that merge at $E=0$, thus making the system metallic.
In the presence of an orthogonal magnetic field, these bands transform into completely flat LLs, of energy reported in \ffref{fig:AB_AA}(c,d).
These LLs are calculated for large bilayer graphene ribbons of width 150~nm, i.e., larger than the magnetic length $\ell\approx 25 / \sqrt{B[{\rm T}]} \ {\rm nm}$ for high enough magnetic fields, to avoid limitation related to the quantized flux through the 2D cell. 
Additional details are reported in \aref{sec:appendixC} and \aref{sec:appendixD}.
Bilayer graphene with AB stacking shows a LL dependence on the magnetic field in between a Dirac and a parabolic dispersion, as predicted by using a continuous Hamiltonian~\cite{McCann2006}, although with different values.
In particular, LLs($n$) with $n\in\mathbb{Z}$ can be roughly approximated, as illustrated by the continuous lines in \fref{fig:AB_AA}(c), by
\begin{equation} \label{eq:LLort}
 {\rm LL}(n) \ \approx \ \left\{ \begin{array}{ll} 
										 0 \ \ \ {\rm and} \ \ \ 1.08 \ \sqrt{B[{\rm T}]}~{\rm meV}  & {\rm if}~n=0 \\[3mm]
	                   352~{\rm meV} \ + \ 2.8 \ n^{1.2} \ B[{\rm T}]^{0.85}~{\rm meV} & {\rm if}~n>0 \\[3mm]
	                   14.3 \ n^{0.59} \ B[{\rm T}]^{0.65}~{\rm meV} & {\rm if}~n>0 \\[3mm]
	                   -13.2 \ |n|^{0.57} \ B[{\rm T}]^{0.65}~{\rm meV} & {\rm if}~n<0 \\[3mm]
	                   -352~{\rm meV} \ - \ 3.9 \ |n|^{1.2} \ B[{\rm T}]^{0.8}~{\rm meV} & {\rm if}~n<0									
									\end{array} \right.
\end{equation}
In a bilayer graphene with AA stacking, as illustrated by the continuous lines in \fref{fig:AB_AA}(d), the LLs dispersion is analogous to that of a monolayer graphene, except for an energy shift
\begin{equation}
  {\rm LL}(n) \ \approx \ \left\{ \begin{array}{ll} 
	                   352~{\rm meV} \ \pm \ 35 \ \sqrt{|n|} \ \sqrt{B[{\rm T}]}~{\rm meV} \\[3mm]
										-352~{\rm meV} \ \pm \ 37 \ \sqrt{|n|} \ \sqrt{B[{\rm T}]}~{\rm meV}							
									\end{array} \right.
\end{equation}

\begin{figure}[t!]
	\centering	
		\includegraphics[width = 8 cm]{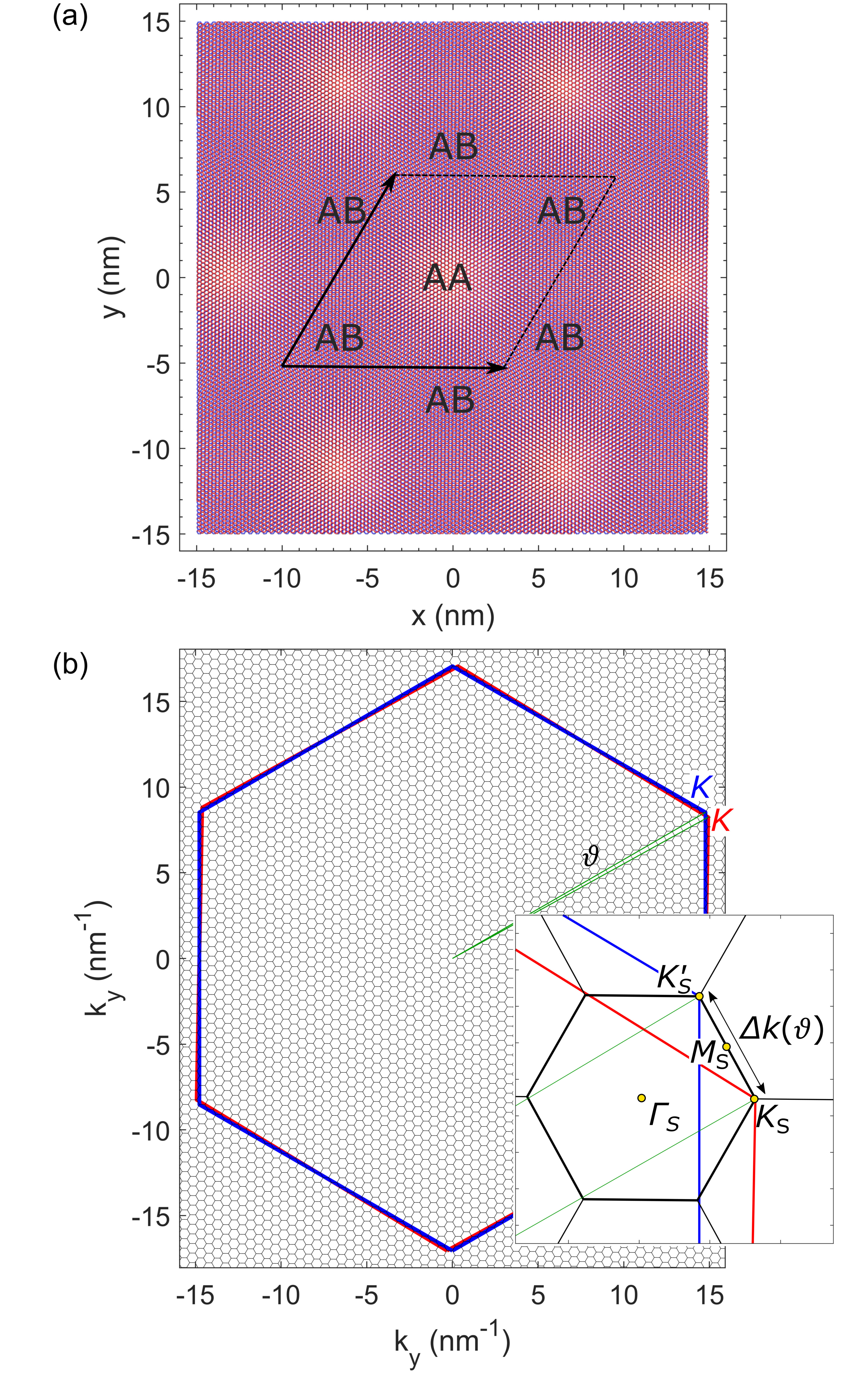}
	\caption{
	   (a) Geometry of the TBG with the magic twist angle $\theta=1.08^\circ$. 
	    	 The unit cell is indicated along with the translation vectors. 
				 The labels AA and AB correspond to the regions where the stacking is approximately of AA or AB type. 
	   (b) Brillouin zones of bottom (blue line) and top (red line) monolayers, with indication of their $K$ points. 
				 The black line corresponds to the reciprocal lattice of the TBG, with a smaller Brillouin zone. 
		     The inset shows the smaller Brillouin zone of the TBG and its $K_{\rm S}$, $K'_{\rm S}$, $M_{\rm S}$ and $\Gamma_{\rm S}$ points.
				$\Delta(\theta)$ is the distance between the $K_{\rm S}$ and $K'_{\rm S}$ points and it corresponds to the distance between the $K$ points of the twisted monolayers as reported in \eref{eq:dk_theta}.
				}
	\label{fig:twisted}
\end{figure}

\begin{figure}[t!]
	\centering	
		\includegraphics[width = 16 cm]{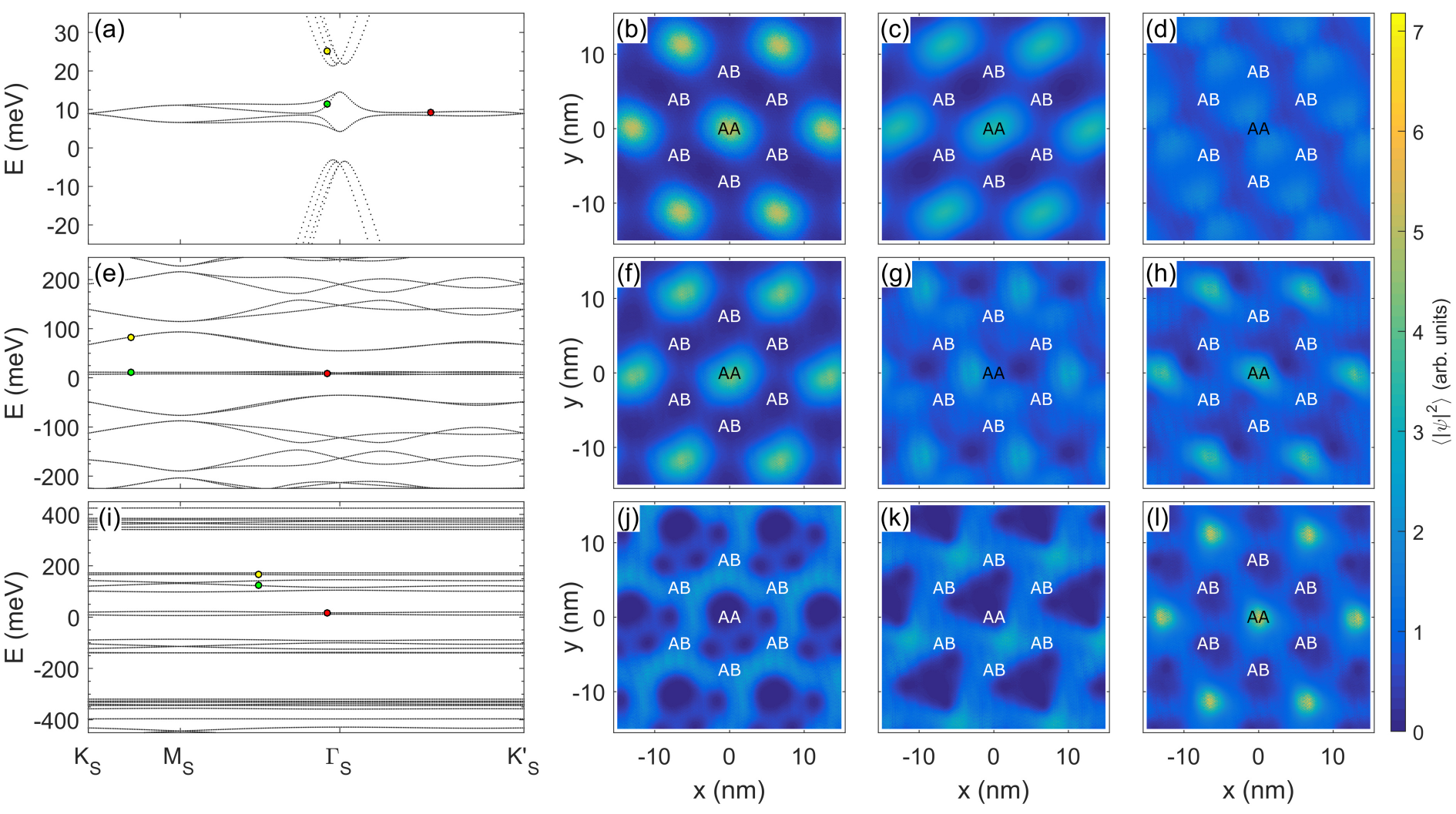}
	\caption{
	   (a) Energy bands for the TBG with twist angle $\theta=1.08^\circ$ in the absence of a magnetic field. 
		 (b,c,d) Spatial density probability averaged over a radius of 0.2~nm of the eigenstates corresponding to the red, green and yellow dots in (a), respectively. 
		  The labels indicate the position of the AA and AB regions, as in \fref{fig:twisted}(a).
	   (e,f,g,h) Same as (a,b,c,d) in the presence of an orthogonal magnetic field $B\approx 28.36$~T, corresponding to the first vertical line in \fref{fig:AB_AA}(c,d).
		 (i,j,k,l) Same as (a,b,c,d) in the presence of an orthogonal magnetic field $B\approx 141.80$~T, corresponding to the second vertical line in \fref{fig:AB_AA}(c,d).
		}
	\label{fig:B_perp}
\end{figure}

\subsection{Orthogonal magnetic field in twisted bilayer graphene\label{sub:ORT}}
For the TBG, we consider the magic twist angle $\theta=1.08^\circ$, which corresponds to a unit cell of area 148.8~nm$^2$, with 11,164 atoms, see \fref{fig:twisted}(a). 
This imposes an \mbox{out-of-plane} component of the magnetic field equal to multiples of 28.36~T. 
In the absence of a magnetic field, the energy bands along the \mbox{$K_{\rm S}-M_{\rm S}-\Gamma_{\rm S}-K'_{\rm S}$} path in the Brillouin zone, see the inset of \fref{fig:twisted}(b), show the expected almost flat \mbox{low-energy} bands within a gap of about 25 meV~\cite{Lin2018}, see \fref{fig:B_perp}(a). 
The charge neutrality point is not exactly at $E=0$, except for the monolayer graphene, due to the calibration of the TB model, and can slightly vary depending on the magnetic field.
The eigenstates of the almost flat bands are mostly localized on the AA stacked subregions~\cite{Laissardiere2012}, with a small probability density on the subregions with intermediate stacking, see \ffref{fig:B_perp}(b,c,d).
At the smallest possible \mbox{non-zero} orthogonal magnetic field of 28.36~T, see \fref{fig:B_perp}(e), almost flat very \mbox{low-energy} bands are still present and the other LLs are dispersive. 
This is the consequence of the local variation of the band dispersion from \mbox{parabolic-like} to \mbox{Dirac-like} in subregions of different stacking. 
A direct comparison of these results with the Hofstadter butterflies obtained with a continuous Hamiltonian~\cite{Bistritzer2011} is not straightforward, because of the different cell sizes and our limited access to magnetic fields corresponding to integers of $\Phi_0$. 
However, the band structure is qualitatively similar to the one obtained in \cref{HerzogArbeitman2022} for conditions analogous to ours.
According to \ffref{fig:AB_AA}(c,d), we should expect the \mbox{low-energy} states to be mainly localized in the AB subregions, and the slightly \mbox{higher-energy} states mainly localized in the AA subregions.
However, \ffref{fig:B_perp}(f,g,h) show that this is not the case. 
\mbox{Low-energy} states are still confined in the AA subregions, as in the absence of a magnetic field.
This is because the magnetic length $\ell\approx 4.88$~nm is comparable to the size of the AA and AB subregions, and thus Landau states cannot fully develop inside them. 
Moreover, the energy proximity of LLs in the AA and AB subregions, shown by the first vertical line in \ffref{fig:AB_AA}(c,d), facilitates their mixing for TBGs. 
For a larger magnetic field, the LLs turn flatter and flatter, because of the shorter magnetic length. 
\Fref{fig:B_perp}(i) shows the energy bands at $B=141.80$~T, displayed by the second vertical line in \ffref{fig:AB_AA}(c,d). 
This time, thanks to the short magnetic length, $\ell\approx 2.18$~nm, the \mbox{low-energy} states are mainly confined in the AB subregions, corresponding to LL($n=0$) in \fref{fig:AB_AA}(c), see \ffref{fig:B_perp}(j,k,l). 
On the first LL, as expected, the states with flat distribution are mainly confined in the AA subregions according to \fref{fig:AB_AA}(d). 
They become more slightly delocalized when they acquire velocity or completely delocalized for \mbox{higher-energy} bands.

\subsection{In-plane magnetic field in twisted bilayer graphene\label{sub:PLANE}}
\begin{figure}[b!]
	\centering	
		\includegraphics[width = 16 cm]{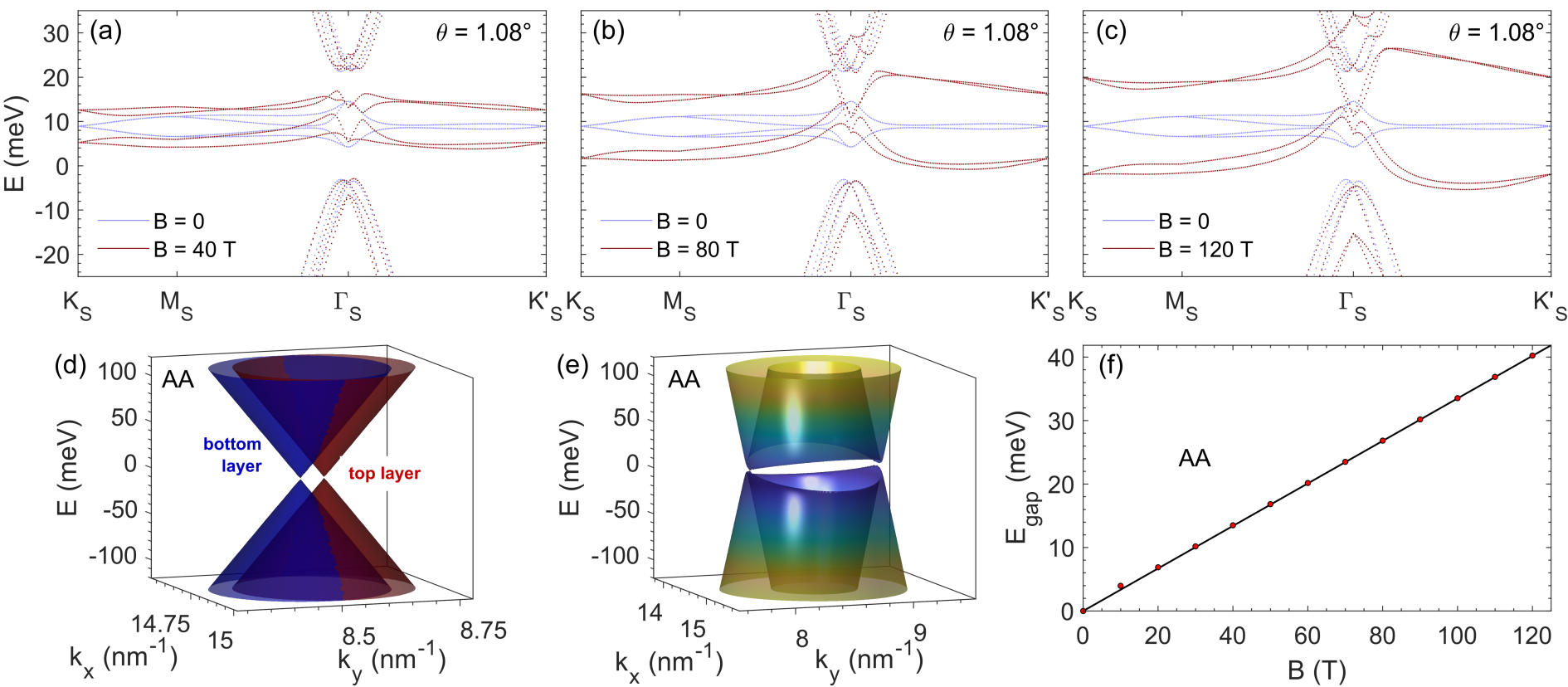}
	\caption{
	    Energy bands of the TBG with twist angle $\theta=1.08^\circ$ in an \mbox{in-plane} magnetic field (a) $B=40$~T, (b) $B=80$~T and (c) $B=120$~T (red dots) compared to the case $B=0$ (blue dots).
	    (d) Dirac cones for uncoupled AA bilayer graphene around the $K$ point and under an \mbox{in-plane} magnetic field $B=120$~T.
			(e) Opening of a gap in the energy bands around the $K$ point for AA bilayer graphene.
			(f) Linear dependence of the maximum value of this energy gap as a function of the \mbox{in-plane} magnetic field strength.
			}
	\label{fig:B_inplane}
\end{figure}
An \mbox{in-plane} magnetic field is not expected to significantly affect the TBG band structure, because the magnetic flux through the thickness of the system is very small, due to the short \mbox{inter-layer} distance $\Delta z$. 
Also, the magnetic flux through each individual layer is zero. 
Indeed, an \mbox{in-plane} magnetic field does not modify the electronic structure of monolayer graphene at all. 
However, from the examples of \ffref{fig:B_inplane}(a,b,c) at $B=40$~T, 80~T and 120~T, it is clear that, at the small magic angle $\theta=1.08^\circ$, the \mbox{low-energy} bands are progressively shifted and modified, and the gap finally closes.
To explain this behavior, we go back to the minimal coupling of \eref{eq:minimal_coupling}, which relates the momentum to the vector potential.
Since the magnetic field is in the \mbox{$xy$-plane}, i.e., $\mathbf{B}=(B_x,B_y,0)$, the corresponding \mbox{in-plane} local shift $\Delta k$ of the wave vectors is orthogonal to the magnetic field and depends on the $z$ coordinate
\begin{equation} \label{eq:kB}
  \Delta\mathbf{k}(\mathbf{B},z) \ = \ \Frac{e}{\hbar c} \ \mathbf{A} \ = \ \Frac{e}{2 \hbar c} \ \mathbf{B}\times  \mathbf{r} 
	                               \ = \ \Frac{e}{2 \hbar c} \ \left( \begin{array}{c}  B_y \\ -B_x\end{array}\right) \ z \ .
\end{equation}
In \eref{eq:kB}, we considered the \mbox{so-called} symmetric gauge for the vector potential, because it has a simple form that allows us to easily interpret the results. 
Any other gauge would simply make the results of the calculations harder to analyze, while leading to the exact same conclusions illustrated below, due to the invariance of the physical observables under gauge transformations. 
The relative shift of the Dirac points in the isolated top and bottom layers is given by
\begin{equation} \label{eq:kBBz}
  \Delta\mathbf{k}(\mathbf{B}) \ \equiv  \ \Delta\mathbf{k}(\mathbf{B},\Delta z)   \    =    \ \Frac{e}{2 \hbar c} \ \left( \begin{array}{c}  B_y \\ -B_x\end{array}\right) \ \Delta z    
\end{equation}
and then
\begin{equation} \label{eq:kBz}
  	\Delta k(B)   \    =    \  \left| \Delta\mathbf{k}(\mathbf{B}) \right|   \ \approx \ 2.5\times 10^{-4} \ B[{\rm T}] \ {\rm nm}^{-1} \  . 
\end{equation}
This equation is detailed in~\cref{Donck2016}, where the analysis is performed on an \mbox{AB-stacked} bilayer graphene with a continuous Hamiltonian and related to misaligned graphene stacking~\cite{He2014,Daboussi2014}, which induces a pseudomagnetic field that still preserves \mbox{time-reversal} symmetry. 
A similar analysis for TBGs is presented in~\cite{Kwan2020}, still with a continuous Hamiltonian limited to the almost flat bands and a relatively small \mbox{in-plane} magnetic field. 
The \mbox{magnetic-field} induced shift of \eref{eq:kBz} acts as spatial shift between the reciprocal spaces of the bottom and top layers and not as a rotation. 
Therefore, it does not affect the system periodicity.
From \eref{eq:kBz}, $\Delta k(B)$ turns out to be very small and should be compared to the shift of the $K$ points in TBGs~\cite{Santos2007} to be relevant
\begin{equation} \label{eq:dk_theta}
  \Delta k(\theta) \ = \  2 \ \left| \mathbf{K} \right| \ \sin(\theta/2) \ \approx \ 17 \ \theta[{\rm rad}] \ {\rm nm}^{-1} \ ,
\end{equation}
where $\left| \mathbf{K} \right|$ is the distance between the $\Gamma$ and $K$ points in the Brillouin zone of a monolayer graphene and the approximation is for small angles, see  \fref{fig:twisted}(b).
For general twist angles, it is exactly this shift $\Delta k(\theta)$ that was found to be the origin of observed van Hove singularities~\cite{Li2009} and begun to attract the interest toward TBGs.
It follows that the ratio
\begin{equation}
  \Frac{\Delta k(B)}{\Delta k(\theta)} \ \approx1.5 \times  10^{-5} \ \Frac{B[{\rm T}]}{\theta[{\rm rad}]} 
\end{equation}
is significant only for small angles $\theta$ and a large \mbox{in-plane} magnetic field $B$.
The combination of the \mbox{Dirac-points} shift and the \mbox{inter-layer} coupling explains the effect observed at low energy for the small magic angle $\theta=1.08^\circ$. 
An even larger effect is expected for $\theta=0$, which corresponds to the AA stacking. 
In \fref{fig:B_inplane}(d), we see the shift $\Delta k(B)$ induced by an \mbox{in-plane} magnetic field $B=120$~T along the \mbox{$x$-direction} on the Dirac points of the two isolated monolayers that compose an AA bilayer graphene. 
As expected, the effect is larger along the \mbox{$y$-direction}, which is orthogonal to the magnetic field.
\Fref{fig:B_inplane}(e) shows the energy bands around the $K$ point of an AA bilayer graphene, with the opening of an energy gap along the \mbox{$y$-direction}.
Such a gap has a maximum value that is proportional to the \mbox{in-plane} magnetic field, as shown in \fref{fig:B_inplane}(f), where $E_{\rm gap}\approx 0.335 \ B[{\rm T}]$~meV, which is therefore observable at relatively low magnetic fields. 
Remarkably, we observe a transition from metallic to semimetallic phases, with possible experimental implications. 
We verified that the angle the magnetic field forms with the \mbox{$x$-axis} does not affect our results and conclusions. 
This is a consequence of the symmetry of the band structure around the Dirac points at low energies. 
In the different situation of low magnetic fields for superconducting TBG, an anisotropy is observed as a consequence of the nematicity of superconductive states~\cite{Cao2021a,Yu2021}. 
While this is very interesting, here we consider the opposite regime where electron interactions do not play a central role, and stronger magnetic fields, where TBGs are in the normal state. 
  
\section{Conclusions}
We demonstrated the importance of the intensity of an orthogonal magnetic field on TBG. 
For a relatively small magnetic field, yet up to tens of T, we observed dispersive bands for the magic angle $\theta=1.08^\circ$. 
When increasing the magnetic field, thus decreasing the magnetic length, flatter LLs appeared, of nature confirmed by the spatial density of the eigenstates.
In the presence of a high \mbox{in-plane} magnetic field, we showed an important effect on the energy gap of the TBG. 
In case of AA stacking, a metallic to semimetallic transition was observed, despite the null magnetic flux through the unit cell. 
This effect is the direct consequence of the minimal coupling, which entails a band shift in the reciprocal space depending on the top or bottom layer. 
We expect such an effect to be even more relevant for multilayer graphene or van der Waals homo- or heterostructures in general. 
Though not taking into account the \mbox{electron-electron} interaction, which is more relevant for \mbox{high-quality} samples and low temperatures, our results should be experimentally observable by measuring electronic transport~\cite{Arrighi2023} or density of states, for example using scanning tunneling microscopy~\cite{Liu2023}.

\appendix

\section{Choice of the tight-binding model}
The use of TB models allows a spatial microscopic description of the bilayer graphene. 
Among the many TB models with a single $p_z$ orbital {\it per} carbon atom and existing in the literature, the one in \cref{Laissardiere2012} is the first one able to reproduce, without any geometrical relaxation, almost flat \mbox{low-energy} bands expected for magic angles. 
However, the energy gap is not observed.  
Such a \mbox{Slater-Koster} model~\cite{Slater1954}, based on \mbox{intra-layer} and \mbox{inter-layer} Hamiltonian elements depending on the \mbox{carbon-carbon} distance with a \mbox{cut-off} distance of 0.6~nm, was used to clearly demonstrate that the states in the almost flat bands are localized in the AA stacked regions.

The more complex model of \cref{Fang2016} takes into account the relative positions of neighboring atoms by considering the dependence of the coupling element between two atoms on the angles that they form with neighboring atoms. 
This introduces an environmental dependence, which is useful to more accurately describe strained systems.
Again, this model correctly reproduces the nearly flat \mbox{low-energy} bands, where the states are mainly concentrated in the AA regions, this time with a small energy gap just above.

A more accurate local environment TB model is reported in \cref{Pathak2022} and based on the \mbox{environment-dependent} TB model of \cref{Tang1996}.
It reproduces, for magic angles, the almost flat \mbox{low-energy} bands within a bandgap and can easily take into account strain and dislocations in the bilayer graphene, which is important if we consider the geometrical relaxation.
    
Relaxing the system geometry can be difficult and numerically heavy if the twist angle is small, thus forming a large primitive cell. 
Since we are not interested here in relaxing the system, we use the TB model of \cref{Lin2018}, which reproduces the almost flat bands within a small gap for the magic angle $\theta=1.08$°.
It also well reproduces the \mbox{density-functional} energy bands for AA and AB stackings, and in particular the shifted \mbox{parabolic-like} and \mbox{Dirac-like} bands observed and discussed in \cref{Laissardiere2012}.

\section{Details about the Peierls phase}
In the case of a periodic \mbox{two-dimensional} system, the Peierls phase detailed in \cref{Cresti2021} preserves the spatial periodicity of the system, thus allowing the use of the Bloch theorem to calculate the energy bands.
We indicate the lattice vectors of the \mbox{two-dimensional} periodic system as $\mathbf{T}_1$ and $\mathbf{T}_2$, while the position of the carbon atom inside the unit cell is given by the basis vectors $\{\mathbf{d}_i\}$. 
If the monolayers that compose the TBG lie on to the \mbox{$xy$-plane}, then the \mbox{$z$-components} of the lattice vectors vanish, i.e., $\mathbf{T}_1=\left(T_1^x,T_1^y,0\right)$ and $\mathbf{T}_2=\left(T_2^x,T_2^y,0\right)$, while the \mbox{$z$-components} of the basis vectors are equal to $\pm\Delta z/2$, i.e., $\mathbf{d}_i=\left(d_i^x,d_i^y,d_i^z=\pm\Delta z/2\right)$, if the atoms are on the top layer or bottom layer, respectively. 
According to \cref{Cresti2021}, in the presence of a magnetic field $\mathbf{B}=\nabla\times\mathbf{A}$, where $\mathbf{A}$ is the vector potential, the Hamiltonian element $H_{im_1n_2,jm_2n_2}$ between the $p_z$ orbitals of the carbon atoms at position $R_{im_1n_1}=d_i + m_1 \mathbf{T}_1 + n_1 \mathbf{T}_2 $ and $R_{jm_2n_2}=d_j + m_2 \mathbf{T}_1 + n_2 \mathbf{T}_2 $, is multiplied by the factor $\exp(i\varphi_{i,m_1,n_1;j,m_2,n_2})$, with $\varphi_{i,m_1,n_1;j,m_2,n_2}$ the Peierls phase given by
\begin{equation} \label{eq:peierls} 
	\varphi_{i,m_1,n_1;j,m_2,n_2}  =  
				\Frac{\pi}{\Phi_0} \ \mathbf{B} \cdot
				\left[ 
				         \mathbf{d}_i \times \mathbf{d}_j + 
								 (\mathbf{d}_i + \mathbf{d}_j ) \times \left[  (m_2-m_1) \mathbf{T}_1 -  (n_2-n_1) \mathbf{T}_2 \right] 
								 + (m_1+m_2)(n_2-n_1) \mathbf{T}_1 \times \mathbf{T}_2 
                                         \right]
	\ , 
\end{equation}
where, in order to keep the spatial periodicity, the flux of the magnetic field through the unit cell must be a multiple of the quantum flux $\Phi_0$, i.e., 
\begin{equation} \label{eq:flux}
	\mathbf{B}\cdot\left( \mathbf{T}_1 \times \mathbf{T}_2 \right) \ = \ q \ \Phi_0 \ \ \ {\rm with} \ \ \ q\in\mathbb{Z} \ . 
\end{equation}
If the magnetic field is orthogonal to the bilayer graphene and the oriented along $z$, i.e., $\mathbf{B}=(0,0,B_z)$, then we have
\begin{eqnarray} \label{eq:peierls_orthogonal} 
	\varphi_{i,m_1,n_1;j,m_2,n_2} = \Frac{\pi}{\Phi_0} B_z \!
				\left[ \begin{array}{l}
								d_i^x d_j^y - d_i^y d_j^x +
							( d_i^x + d_j^x) \left[ (m_2 - m_1 ) T_1^y + (n_2 - n_1 ) T_2^y \right] + \\[3mm]
							+  ( d_i^y + d_j^y) \left[ (m_2 - m_1 ) T_1^x + (n_2 - n_1 ) T_2^x \right] 
							+ T_1^x T_2^y - T_1^y T_2^x
							\end{array}
			  \right]
	\ ,
\end{eqnarray}
where $B_z$ must be a multiple of the minimum magnetic field, according to \eref{eq:flux}, and the Peierls phase is independent of the \mbox{$z$-position} of the layers.
If the magnetic field is \mbox{in-plane}, i.e., $\mathbf{B}=(B_x,B_y,0)$, then we have
\begin{eqnarray} \label{eq:peierls_in-plane}
	\varphi_{i,m_1,n_1;,j,m_2,n_2} 
	         & = & \Frac{\pi}{\Phi_0} \mathbf{B}\cdot\left\{ \mathbf{d}_i\times\mathbf{d}_j  +  
					            \left(\mathbf{d}_i  +  \mathbf{d}_j \right) \times
											       \left[\left(m_2 -  m_1\right)\mathbf{T}_1  -  \left(n_2 -  n_1\right)\mathbf{T}_2  \right] \right\} 
	\\[3mm] \nonumber  & = & \Frac{\pi}{\Phi_0} \left[ \begin{array}{l}
	                                       B^x  \left\{    d_i^y  d_j^z - d_i^z  d_j^y -  \left( d_i^z  +  d_j^z \right)  
												\left[\left(m_2 -  m_1\right)  T_1^y  -  \left(n_2 -  n_1\right)  T_2^y  \right]
																	\right\}  +  \\[4mm]
												+  B^y  \left\{  d_i^z  d_j^x - d_i^x d_j^z   + 
											    \left( d_i^z + d_j^z \right) 
										\left[\left(m_2  -  m_1\right)  T_1^x  -  \left(n_2 -  n_1\right) T_2^x  \right]\right\}
	                                   \end{array}\right]
	\ , 
\end{eqnarray}
with no limitation on $|\mathbf{B}|$, since the flux through the unit cell is null. 
If we consider the Peierls phase between orbitals of atoms that belong to the same top or bottom monolayer graphene, then $d_j^z=\pm\Delta z/2$ and
\begin{equation}
	\varphi_{i,m_1,n_1;j,m_2,n_2}^\pm 
	          =  \pm \Frac{\pi}{\Phi_0} \Frac{\Delta z}{2}  \left[ \begin{array}{l}
	                                       B^x \left\{ d_i^y  -  d_j^y  -       2 \left(m_2 -  m_1\right)  T_1^y  +  2 \left(n_2 -  n_1\right)  T_2^y \right\}  + \\[4mm]
				     +  B^y  \left\{  d_j^x  -  d_i^x  +  2 \left(m_2  -  m_1\right)  T_1^x  -  2  \left(n_2 -  n_1\right)  T_2^x \right\}
	                 \end{array}\right]
	\ ,
\end{equation}
where the $\pm$ index indicates the top (+) or bottom (-) layer. 
Since the flux of the \mbox{in-plane} magnetic field through each individual layer is null, we expect that the circulation of the Peierls phase along any closed \mbox{in-plane} path vanishes too. 
If the closed path is composed of $N$ atomic positions identified by $\{i_k,m_k,n_k\}$ with $n=1...N$ and $i_N=i_1$, $m_N=m_1$ and $n_N=n_1$, we then have 
\begin{eqnarray}
	\sum_{k=1}^N  \varphi_{i_k,m_k,n_k;i_{k+1},m_{k+1},n_{k+1}}^\pm 
	  & = &
	\nonumber  
         \pm  \Frac{\pi}{\Phi_0}\ \Frac{\Delta z}{2}  \sum_{k=1}^{N-1}  \left[ \begin{array}{l}
	                                       B^x  \left\{ d_{i_k}^y  -  d_{i_{k+1}}^y   -
													              2 \left(m_{k+1} -  m_k\right)  T_1^y  +  2 \left(n_{k+1} -  n_k\right)  T_2^y \right\}  +  \\[4mm]
																				 + B^y  \left\{   d_{i_{k+1}}^x - d_{i_k}^x   + 
																				    2  \left(m_{k+1}  -  m_k\right)  T_1^x  -  2  \left(n_{k+1} -  n_k\right)  T_2^x \right\}
	                                   \end{array}\right] 
		\nonumber \\[3mm]  \hspace{-1cm} & =&  \pm  \Frac{\pi}{\Phi_0} \Frac{\Delta z}{2}  \left[ \begin{array}{l}
	                                       B^x  \left\{ d_{i_1}^y  -  d_{i_N}^y  - 
													              2 \left(m_N -  m_1\right)  T_1^y  +  2 \left(n_N - n_1\right)  T_2^y \right\}  +  \\[4mm]
																			+  B^y  \left\{ d_{i_N}^x  -  d_{i_1}^x   + 
																				    2  \left(m_N  -  m_1\right)  T_1^x  -  2  \left(n_N -  n_1\right)  T_2^x \right\}
	                                   \end{array}\right] 
         \ = \  0 \ . 
\end{eqnarray}
Therefore, on each individual layer, the \mbox{in-plane} magnetic field just acts as a local gauge that shifts the reciprocal space, as expected from the minimal coupling and as shown in \sref{sub:PLANE}.

\section{Bands in AB and AA stacked bilayer graphene and TBG with $\theta\approx 1.08^\circ$ under orthogonal magnetic field \label{sec:appendixC}}
\begin{figure}[ht!]
	\centering	
		\includegraphics[width = 15 cm]{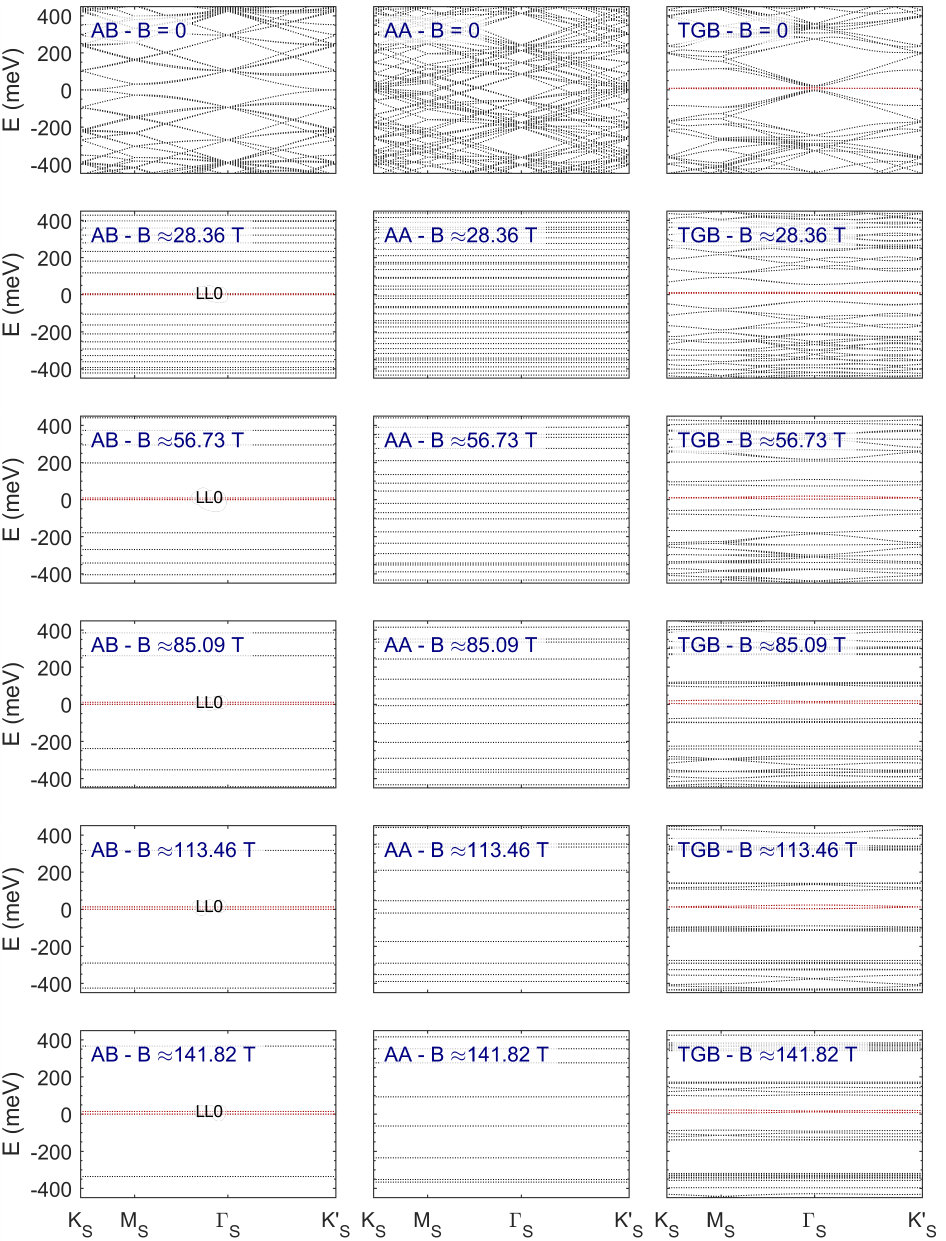}
	\caption{
	   Energy bands for AB and AA stacked bilayer graphenes and TBG with $\theta\approx 1.08^\circ$ under different orthogonal magnetic field intensities $B$, as indicated by the labels.
		In the left panels, the red bands labeled by LL0 correspond to the zero Landau levels, i.e., LL($n=0$) in \eref{eq:LLort}. 
		In the right panels, the red bands are the dispersive \mbox{low-energy} flat bands, of nature, as discussed in \sref{sub:AA_AB}, similar to LL0 only at high magnetic fields.
		}
	\label{S_fig:B_perp}
\end{figure}

\Fref{S_fig:B_perp} reports the energy band structure under orthogonal magnetic field along the path \mbox{$K_{\rm S}$-$M_{\rm S}$-$\Gamma_{\rm S}$-$K'_{\rm S}$} of the Brillouin zone for the AB and AA stacked bilayer graphenes and TBG with $\theta\approx 1.08^\circ$, see \fref{fig:AB_AA}. 
The magnetic field is a multiple of $B\approx 28.36$~T, i.e., the minimum magnetic field of flux through the unit cell equal to the quantum flux $\Phi_0$, see \fref{fig:twisted}(a).
In the presence of a magnetic field, the LLs are flat for the AB and AA stacked bilayer graphenes, while they are dispersive for TBG, as discussed in \sref{sub:ORT}.

\newpage 

\section{Landau levels for AB and AA stacked bilayer graphene ribbons\label{sec:appendixD}}
\begin{figure}[h!]
	\centering	
		\includegraphics[width = 15 cm]{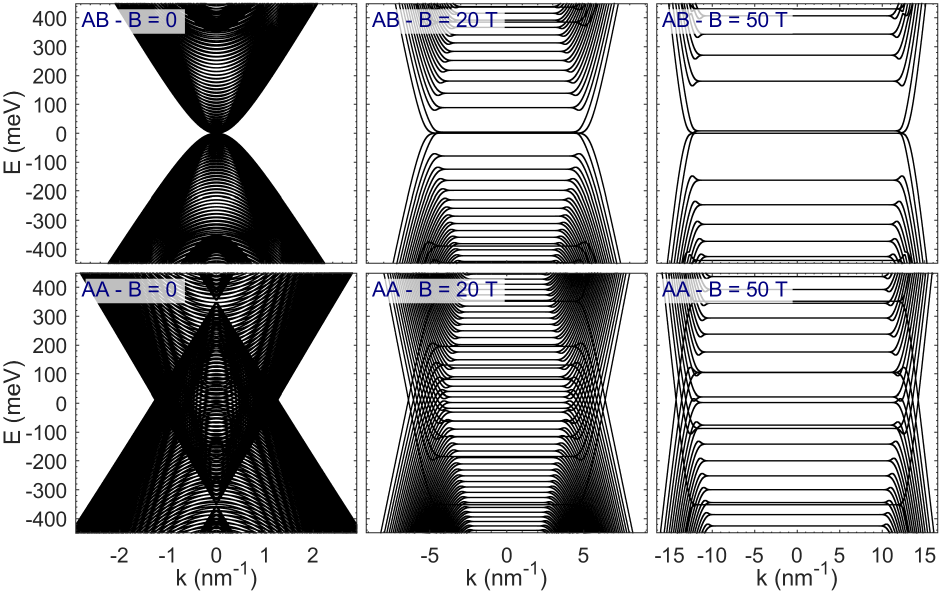}
	\caption{
	   Energy bands of AB and AA stacked bilayer graphene armchair ribbons of width $W\approx 150$~nm and under orthogonal magnetic fields $B=0$, 20~T and 50~T, as indicated by the labels.
		}
	\label{S_fig:ribbon_B_perp}
\end{figure}

To avoid limitations when choosing the magnetic field, we calculate the energy bands of untwisted bilayer graphene armchair ribbon (with AA or AB stacking) of width $W\approx 150$~nm. 
Indeed, the \mbox{one-dimensional} periodicity of the system allows us to choose a gauge that preserves the Hamiltonian periodicity without needing a magnetic flux multiple of $\Phi_0$~\cite{Cresti2021}.
The primitive cell of AB and AA stacked bilayer graphenes contains 4,884 carbon atoms and the lattice vector has a length of about 0.4254~nm. 
The armchair edge geometry has been chosen in order to avoid edge states, which are clearly absent in \mbox{two-dimensional} systems. 
By comparing the ribbon width to the magnetic length, we have 
\begin{equation}
	\Frac{W}{\ell}\approx \Frac{W}{25\ {\rm nm}} \  \sqrt{B[{\rm T}]} \ = \ 6 \ \sqrt{B[{\rm T}]} \ .
\end{equation}
Thus, for $B>10$~T, it follows $W/\ell>18$ and the states in the middle of the ribbon are expected not to be significantly affected by the ribbon edges.
The bands at different magnetic fields for AB and AA stacked bilayer graphene ribbons are reported in \fref{S_fig:ribbon_B_perp}. 
\FFref{fig:AB_AA}(c,d), corresponding to the flat LLs, were obtained this way.
The states with dispersive energy between the LLs correspond to the edge states. 
At $B=0$, the system corresponds to that of \ffref{fig:AB_AA}(a,b). 
However, due to the large but finite width of the ribbon, a small gap appears at the shifted Dirac points in the AA stacked ribbon, see the corresponding panel in \fref{S_fig:ribbon_B_perp}.

\section{Effect of the in-plane magnetic field orientation for AA stacked bilayer graphene}
\begin{figure}[h!]
	\centering	
		\includegraphics[width = 12 cm]{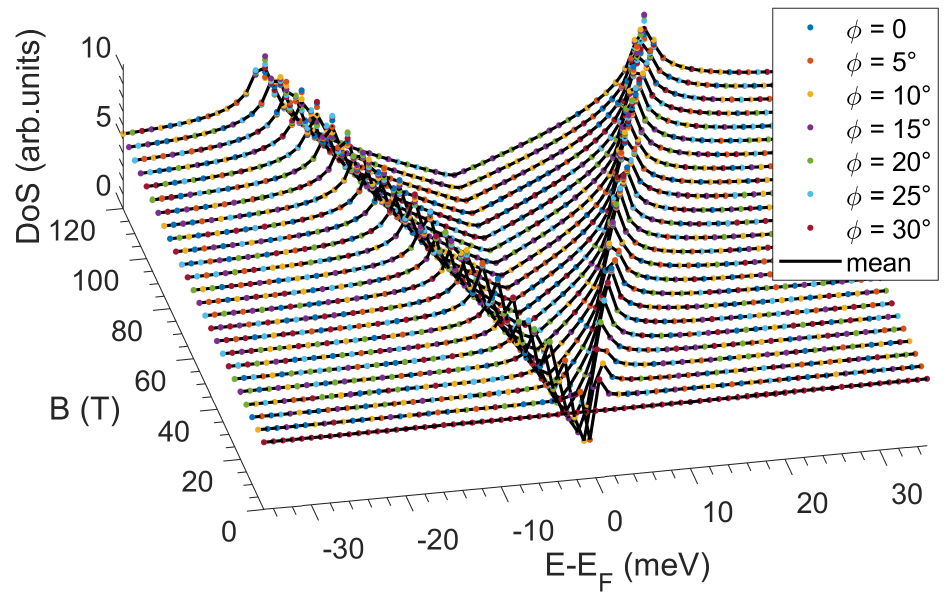}
	\caption{
	          Density of states as a function of energy and magnetic field intensity, for an AA stacked bilayer graphene. 
						The dots correspond to different angles $\phi$ of the \mbox{in-plane} magnetic field with the \mbox{$x$-axis}. 
						The black lines correspond to the average density of states.
		}
	\label{S_fig:AA_B_parallel}
\end{figure}

The Dirac bands of monolayer graphene can be considered as perfectly symmetric cones at low energies. 
Therefore, in TBGs, changing the angles $\phi$ of an \mbox{in-plane} magnetic field with the \mbox{$x$-axis} does not affect the electronic structure in the small energy range of the gap. 
The only difference, not shown here, is the orientations of the bands, which are equally affected by the choice of the gauge. 
\Fref{S_fig:AA_B_parallel} shows that the density of states is independent of the angle $\phi$, except for small numerical variations around the density of states peak near the maximum energy gap.  
These numerical errors are the consequence of the calculation methodology, which directly extracts the density of states by counting the number of \mbox{$k$-points} within a 1~meV range around each energy in the band structure. 
From the density of states, we observe the metallic (for $B=0$) to semimetallic transition, with a linear dependence of the larger energy gap, as detailed in \sref{sub:PLANE}.

\section{Some details about the calculation}
To calculate the eigenvectors and eigenvalues of the Hamiltonian matrices, we used \textsc{Matlab}~\cite{MATLAB}. 
In the case of the magic angle $\theta\approx 1.08^\circ$, as specified in \sref{sub:ORT}, the unit cell contains 11,164 carbon atoms and is connected to the eight neighboring cells. 
The \mbox{$k$-dependent} Hamiltonian matrix is thus a 11,164$\times$11,164 Hermitian matrix. 
Given the TB model, it has 369,026 nonzero elements, which corresponds to a sparsity of about $0.2962\%$. 
Calculating all the eigenvalues would be very long and useless, since we are focusing on a relatively small energy range of about $\pm 500$~meV. 
We then made use of the function
\vspace*{-0.5mm}
\begin{lstlisting}[style=Matlab-editor]
	eigs(h,n,'smallestabs') 
\end{lstlisting}
\vspace*{-0.5mm}
where \lstinline[style=Matlab-editor]!h! is the matrix to diagonalize, while \lstinline[style=Matlab-editor]!n! and \lstinline[style=Matlab-editor]!'smallestabs'! indicate that the first \lstinline[style=Matlab-editor]!n! eigenvalues with smallest absolute value have to be found. 
In our case \lstinline[style=Matlab-editor]!n!=100. 
The calculation is parallelized on the \mbox{$k$-points}.

\end{document}